\documentclass[aps,pre,showpacs,twocolumn,10pt,showkeys]{revtex4-1}
\usepackage{graphics}
\usepackage{amsmath}
\usepackage{color}
\usepackage{verbatim}
\usepackage{siunitx}
\usepackage{multirow}
\usepackage{amssymb}
\usepackage{amsfonts}
\usepackage{amsthm}

\usepackage[normalem]{ulem}	

\begin{document}
\title{Study of Conduction Cooling Effects in Long Aspect Ratio Penning-Malmberg Micro-Traps}
\author{M. A. Khamehchi}
\email{mohammad.khamehchi@email.wsu.edu}
\author{C. J. Baker}
\author{M. H. Weber}
\author{K. G. Lynn}
\email{kgl@wsu.edu}
\affiliation{Department of Physics and Astronomy, Washington State University, Pullman, Washington  99164-2814}

\begin{abstract}
{A first order perturbation with respect to velocity has been employed to find the frictional damping force imposed on a single moving charge due to a perturbative electric field, inside a long circular cylindrical trap. We find that the electric field provides a cooling effect, has a tensorial relationship with the velocity of the charge. A mathematical expression for the tensor field has been derived and numerically estimated. The corresponding drag forces for a charge moving close to the wall in a cylindrical geometry asymptotically approaches the results for a flat surface geometry calculated in the literature. Many particle conduction cooling power dissipation is formulated using the single particle analysis. Also the cooling rate for a weakly interacting ensemble is estimated. It is suggested that a pre-trap section with relatively high electrical resistivity can be employed to cool down low density ensembles of electrons/positrons before being injected into the trap. For a micro-trap with tens of thousands of micro-tubes, hundreds of thousands of particles can be cooled down in each cooling cycle. For example, tens of particles per micro-tube in a $5 cm$ long pre-trap section with the resistivity of $0.46 \Omega m$ and the micro-tubes of radius $50 \mu m$ can be cooled down with the time constant of $106\mu s$.}
\end{abstract}
\maketitle

\section{Introduction}
\begin{figure}[as soon as possible]
\includegraphics{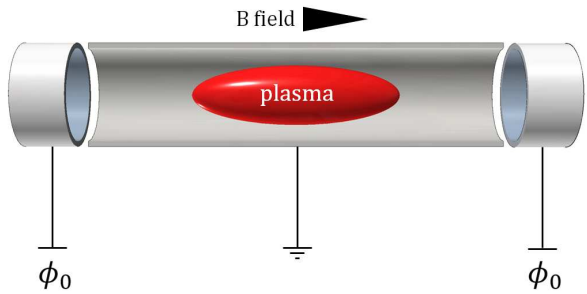}
\caption{\label{fig-1}A schematic sketch of a Penning-Malmberg trap with $\phi_{0}$ as the end cap potential in the constant magnetic field of $B$.}
\end{figure}

Penning-Malmberg (PM) traps \cite{Penning1936873,malmberg1975properties} are widely used for storing non-neutral plasmas because of the ease of construction using a static magnetic field. A schematic sketch of a PM trap biased to a potential of $\phi_{0}$ and placed in the constant magnetic field of magnitude $B$, is shown in Fig. \ref{fig-1}.

Depending on the application, cooling processes are of importance for non-neutral plasmas inside a PM trap which can happen both artificially (see, for example, Ref. ~\onlinecite{wineland_cooling1995}) and naturally. Depending on the application and the configuration of the trap, different cooling methods such as laser cooling \cite{PhysRevLett.57.70} and energy exchange cooling \cite{PhysRevE.71.036406} have been employed as ex-situ cooling in conventional traps. As a major in-situ cooling mechanism for trapping positrons (our plasma of interest), buffer gas cooling has been successfully employed in pre-trapping stages \cite{surko2004emerging}. Plasmas can also cool down via naturally occurring processes such as cyclotron radiation, collision, and conduction. We study the latter in this work.

In traps of interest to the present work, the long aspect ratio PM with a few tens of microns in radius and ten centimeters long (micro-trap\cite{folegati2011positron}), many of the cooling mechanisms may not be effective or not applicable. As mentioned, radiation cooling is one of the dominant naturally occurring in-situ cooling mechanisms in conventional traps that utilizes high magnetic fields. However, in a micro-trap, this mechanism is inhibited due to the high cut frequency of the trap as a waveguide. Therefore, in designing a micro-trap, one needs to take into account other cooling effects which are necessary to reduce the particle loss in the plasma.

Conduction cooling, as a cheap and effective method, can be employed in a micro-trap as a naturally occurring cooling process due to the penetration of the electric velocity fields of the moving charged particles into the wall and creating current that dissipate the energy into ohmic heat in the wall. In some cases, depending on the method employed, loading the trap might require precise temporal control over the trapping potential array which requires more precise information on the forces on the particles and the kinetics inside the trap. Our results show that the cooling force has an inverse cubic relationship with the trap radius and a linear dependence on the wall resistivity. Therefore in a trap with a few tens of microns in radius and large enough electrode resistivity, conduction cooling can be an important effect and should not be ignored.
\subsection{A short history on the drag force on image charges}
The velocity fields, due to moving charged particles near a conductor, cause algebraically decaying field penetration into the conductor which is completely different from the exponentially decaying penetration of the radiation fields \cite{boyer1974penetration}. Velocity field derivations and calculations have had applications in the Bohm-Oharanov effect \cite{Mollenstedt}. Skin-depth based analysis fails to obtain the correct results for the internal fields and the surface charge densities created by a moving charge travelling outside of, and close to, a conductive surface. Unsuccessful skin-depth analysis has been adapted (see, for example, Ref. ~\onlinecite{Ekasper1966}) in order to approximate the field-surface interactions for different purposes \cite{boyer1974penetration,boyer1999understanding}.

First order approximation velocity field analysis for a charge with slow velocity moving in parallel with a flat conductive surface/slab has been done by Boyer \cite{boyer1974penetration,boyer1999understanding,boyer1996penetration}. Jones confirms Boyer's conclusion and shows that in Cherenkov radiation however skin effect is dominant \cite{0305-4470-8-5-009}. The same perturbative method has been employed to find the velocity electric fields due to a charge moving inside a conducting sphere \cite{Aguirregabiria19956}. An analysis has been done by Schaich for a charge moving with arbitrary velocity in parallel with a conductive slab, and the low velocity results confirms Boyer's perturbation analysis \cite{PhysRevE.64.046605}.
\subsection{Positron storage}
During recent years low energy positron sources are becoming increasingly important in many areas of physics and technology, including atomic physics research \cite{Kauppila19891}, plasma physics \cite{greaves:1439}, astrophysics \cite{PhysRevLett.53.2347}, mass spectrometry \cite{HulettJr1993236}, antihydrogen physics \cite{Charlton199465}, and materials research \cite{Krause-Rehberg1999positron,lynn1988interaction}. In many cases tunable positron sources, both in energy and intensity, are necessary and can be achieved by using particle traps \cite{surko2004emerging}. However, there are unavoidable limitations on the number of particles in such devices which has led to the a new design of a multi-cell trap \cite{surko2003multicell,danielson2006plasma,greaves2002practical}.

\begin{figure}
\includegraphics{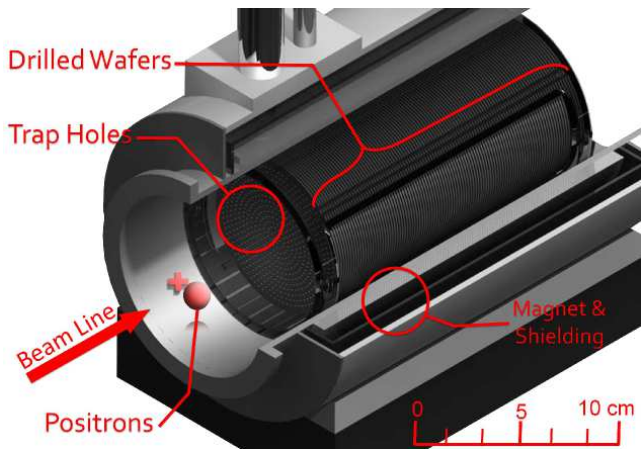}
\caption{\label{micro_trap} The micro-trap concept. Drilled conductor wafers are stacked up to build thousands of parallel long aspect ratio micro-traps.}
\end{figure}

Micro-traps can be employed as proposed by one the authors (K. G. Lynn \cite{lynngreaves2001longaspect}) in order to reduce the minimum confining potential ($\phi_{0}$) at the two end electrodes of a PM trap which can be obtained from
\begin{equation}
\phi_{0}=\frac{eN_{p}}{4\pi\epsilon_{0}L_{p}}\left[2ln\left(\frac{a}{\rho_{p}}\right)+1\right],\label{eq:0.1}
\end{equation}
where $e$ is the positron charge, $N_{p}$ is the number of positrons, $L_{p}$ is the plasma length, $\epsilon_{0}$ is the vacuum permittivity, and $a$ and $\rho_{p}$ are the radius of the trap and the plasma \cite{surko2003multicell}. Thus, for \num{E13} positrons in a trap of length 10 cm and even with $(a/\rho_{p})=1$ which is practically impossible, the confining potential is \num{1.4E5}V which is hard to maintain especially for a portable device. Using micro-traps reduces this amount significantly since the trap walls tend to neutralize the space charge. Recently, simulations have shown the feasibility of making such devices with enhanced total number of particles with a confining potential of a few tens of volts \cite{jialireza2012feasibility}. Figure \ref{micro_trap} shows the concept of a compact positron trap. The long aspect ratio micro-traps are created via stacking up identical etched conductive wafers. The wafers are surrounded by a radiation shield, and a superconductor magnet to provide the static solenoid magnetic field.

Before we start our calculation, it might be useful to mention that the approach in Sections \ref{sec:ii} and \ref{sec:iii} is general and can be applied to any case involving charged particles moving in a long conductive cylinder with or without a magnetic field; however, the thermalization time estimations are performed for positrons/electrons inside a long aspect ratio micro-trap as is our case of interest.
\section{\label{sec:ii}Method and Calculation: Single Particle Analysis}
Using a cylindrical coordinate system of ($\rho,\phi,z$), the electric potential inside a conductive cylinder, with length $L$ and radius $a$ oriented along the $z$ axis, due to a point charge $q$ is given by \cite{jackson1998classical}
\begin{eqnarray}\label{eq:1}
\Phi(\vec{x},\acute{\vec{x}}) &=& \frac{q}{\pi\epsilon_{0}a}\nonumber \\
&&\times\sum\limits_{m=-\infty}^\infty \sum\limits_{n=1}^\infty e^{im(\phi-\acute{\phi})}\frac{J_{m}\left(\frac{x_{mn}\rho}{a} \right) J_{m}\left(\frac{x_{mn}\acute{\rho}}{a} \right)}{x_{mn}J_{m+1}^2(x_{mn})}\nonumber \\
&&\times\left(\frac {\sinh\left(\frac{x_{mn}z_{<}}{a}\right) \sinh\left(\frac{x_{mn}(L-z_{>})}{a}\right)}{\sinh\left(\frac{x_{mn}L}{a}\right)} \right),
\end{eqnarray}
in which $J_{m}$ is the Bessel function of the first kind and $x_{mn}$ is the Bessel function zeros. $\vec{x}$, and $\acute{\vec{x}}$, are the observation point and the position of the charge, respectively. $z_{>}$ is the maximum value and $z_{<}$ the minimum value of the observation and charge $z$ coordinate respectively.
For a long aspect ratio trap, $L \gg a$ (\textit{e. g.} $\frac{L}{a}=\num{2E3}$), we can simplify Eq. (\ref{eq:1}) to obtain
\begin{eqnarray}\label{eq:2}
\Phi(\vec{x},\acute{\vec{x}}) &=& \frac{q}{2\pi\epsilon_{0}a}\sum\limits_{m=-\infty}^\infty \sum\limits_{n=1}^\infty e^{im(\phi-\acute{\phi})}\nonumber \\
&&\times\frac{J_{m}\left(\frac{x_{mn}\rho}{a} \right) J_{m}\left(\frac{x_{mn}\acute{\rho}}{a} \right)}{x_{mn}J_{m+1}^2(x_{mn})} e^{\left(- \frac{x_{mn}(z_{>}-z_{<})}{a}\right)},
\end{eqnarray}
which can then be solved to find the surface charge density on the inner surface of the cylinder as
\begin{eqnarray}\label{eq:3}
\sigma^{(0)}(\phi,z;\acute{\vec{x}}) &=& -\frac{q}{2\pi a^{2}}\sum\limits_{m=-\infty}^\infty \sum\limits_{n=1}^\infty e^{im(\phi-\acute{\phi})}\nonumber \\
&&\times\frac{J_{m}\left(\frac{x_{mn}\acute{\rho}}{a} \right)}{J_{m+1}(x_{mn})} e^{\left(- \frac{x_{mn}(z_{>}-z_{<})}{a}\right)},
\end{eqnarray}
with the superscript in $\sigma^{(0)}$ indicating the order of the expansion with respect to the charged particle velocity. We have used the identity $J_{m-1}(x)+J_{m+1}(x)=(2n/x)J_{m}(x)$ to simplify Eq. (\ref{eq:3}). It can be seen that Eq. (\ref{eq:3}) is essentially the static surface charge density on the inner surface of the cylinder, and it will be used as a basis to derive the resistive electric field due to the charge motion.
\subsection{\label{mo_axis}Charge motion along the cylinder axis}
Let us assume for the moment that the charge is travelling extremely slowly along the axis of the cylinder. In this case the surface charge density, as calculated by Eq. (\ref{eq:3}), has to move with the charge and with the same velocity. Therefore the charges on the wall have to redistribute themselves in order to keep the surface charge density steady (in the point charge reference frame) as the point charge moves along the cylinder.

We will use this approach in order to find a first order correction to the electric field due to the motion of the charge. The velocity electric field inside the conductor is responsible for the rearrangement of the charges such that for a positive(negative) point charge it attracts(repels) sufficient electrons from the bulk(surface) to the surface(bulk) as the point charge travels along the axis. The rearrangement of the charges inside the conductor dissipates the energy of the particle into the conductor via Joule heating.

The motion of the particle along the $z$ axis implies,
\begin{equation}\label{eq:4}
\frac{\partial \sigma}{\partial t}=- c\beta_{z}\frac{\partial \sigma}{\partial z},
\end{equation}
in which $\sigma$ is the surface charge density, $c$ is the speed of light in vacuum $\beta_{z}$ is $\left(\frac{v_{z}}{c}\right)$ and $v_{z}$ is the velocity of the particle in the cylinder along the cylinder axis, $\hat{z}$. Charge conservation implies
\begin{equation}\label{eq:5}
\frac{\partial \sigma}{\partial t}+j_{\rho i}=0,
\end{equation}
in which $j_{\rho i}$ is the radial current density inside the conductor. Assuming an ohmic conductor we will have
\begin{equation}\label{eq:6}
j_{\rho i}=\frac{E_{\rho i}}{\eta},
\end{equation}
where $\eta$ is the conductor resistivity. As we know to the zeroth order, equivalent to the static case, we have the conditions
\begin{subequations}
\label{static} 
\begin{eqnarray}
E_{\rho i}^{(0)}&=&0,\label{eq:7a}
\\
E_{\rho o}^{(0)}&=& \frac{\sigma}{\epsilon_{0}},\label{eq:7b}
\end{eqnarray}
\end{subequations}
for the fields inside and outside of the conductor. Note that the superscript indicates the perturbation order while the subscript $\rho$ indicates a radial component and '$i$' and '$o$' indicate whether the field is inside or outside the conductor respectively.
Equations (\ref{eq:4}), (\ref{eq:5}) and (\ref{eq:6}) can then be combined to obtain
\begin{equation}
E_{\rho i}= c \beta_{z} \eta \frac{\partial \sigma}{\partial z}. \label{eq:8}
\end{equation}
which has a linear behaviour in $\beta_{z}$. Thus, the first order correction to the electric field on the cylinder surface and just inside the conductor is
\begin{equation}
E_{\rho i}^{(1)}=c\beta_{z}\eta\frac{\partial \sigma^{(0)}}{\partial z},\label{eg:9}
\end{equation}
which can be used as a boundary condition to find the unknown coefficients of an electric potential expansion.

By solving the Laplace's equation in cylindrical coordinates using appropriate separation constants, the potential can be expanded as,
\begin{eqnarray}
\Phi_{i}^{(1)}&=&\sum_{m=-\infty}^{\infty}\int_{-\infty}^{\infty}dk e^{ikz}e^{im\phi} \nonumber \\
&&\times\left[A_{m}(k)K_{m}(|k|\rho)+B_{m}(k)I_{m}(|k|\rho)\right], \label{eq:9:1}
\end{eqnarray}
where $I_{m}$ and $K_{m}$ are the modified bessel functions, and the coefficients $A$ and $B$ are defined by the boundary conditions. Since only $K_{m}$ is bounded at infinity, $B_{m}$ has to be zero for the solution inside the conductor. The nonzero coefficients can be found using,
\begin{equation}
E_{\rho i}^{(1)}=-\frac{\partial\Phi_{i}^{(1)}}{\partial\rho}\mid_{\rho=a},\label{eq:9.2}
\end{equation}
and therefore Equations (\ref{eq:3}), (\ref{eg:9}), (\ref{eq:9:1}), and (\ref{eq:9.2}) imply,
\begin{eqnarray}
&&-\sum_{m=-\infty}^{\infty}\int_{-\infty}^{\infty}dk A_{m}(k)|k|\acute{K_{m}}(|ka|)e^{ikz}e^{im\phi} \nonumber \\
&&=\frac{qc\beta_{z}\eta}{2\pi a^{3}}sgn(z)\sum_{m=-\infty}^{\infty}\sum_{n=1}^{\infty}x_{mn}\frac{J_{m}\left(\frac{x_{mn}\acute{\rho}}{a}\right)}{J_{m+1}(x_{mn})}
e^{\frac{-x_{mn}|z|}{a}}e^{im\phi},\label{eq:9.25}
\end{eqnarray}
where $sgn(z)$ is one for the positive, minus one for the negative, and zero for the zero value of $z$. Note that $\acute{z}$ and $\acute{\phi}$ are set to zero for the sake of brevity. they can be included simply by $z\rightarrow z-\acute{z}$ and $\phi\rightarrow \phi-\acute{\phi}$. Considering the following orthogonality relations,
\begin{subequations}
\label{orthos} 
\begin{eqnarray}
&&\int_{0}^{2\pi}d\phi e^{im\phi}e^{-i\acute{m}\phi}=2\pi\delta_{m\acute{m}},\label{eq:9.3a}
\\
&&\int_{-\infty}^{\infty}dz e^{ikz}e^{-i\acute{k}z}=2\pi\delta(k-\acute{k}),\label{eq:9.3b}
\end{eqnarray}
\end{subequations}
and recalling that $K_{m-1}(x)+K_{m+1}(x)=-2\acute{K_{m}(x)}$ we have,
\begin{eqnarray}
A_{m}(k)&=&-\frac{iqc\beta_{z}\eta}{\pi^{2}a}\left[\frac{sgn(k)}{K_{m-1}(|ka|)+K_{m+1}(|ka|)}\right]\nonumber \\
&&\times\sum_{n=1}^{\infty}x_{mn}\frac{J_{m}\left(\frac{x_{mn}\acute{\rho}}{a}\right)}{J_{m+1}(x_{mn})}
\frac{1}{x_{mn}^{2}+(ka)^{2}}.\label{eq:9.4}
\end{eqnarray}

$A_{m}$ defines the perturbative potential inside the conductor, however we need to find the resistive electric field inside the trap. In order to do so, by setting the inside and outside potentials equal on the boundary ($\rho=a$), we can find the potential inside the trap. Therefore by using the general solution of the Laplace's equation in cylindrical coordinates similar to Eq. (\ref{eq:9:1}) and this time accepting only the finite solutions at $\rho=0$, the potential inside the trap can be expanded as
\begin{equation}
\Phi_{o}^{(1)(z)}=\sum_{m=-\infty}^{\infty}\int_{-\infty}^{\infty}dk B_{m}(k)I_{m}(|k|\rho)e^{ikz}e^{im\phi}.\label{eq:9.5}
\end{equation}
Therefore equality of the potentials on the boundary ($\rho=a$) and the orthogonality relations (\ref{eq:9.3a}) and (\ref{eq:9.3b}) imply,
\begin{equation}
B_{m}(k)I_{m}(|k|a)=A_{m}(k)K_{m}(|k|a).\label{eq:9.6}
\end{equation}

Equations (\ref{eq:9.4}), (\ref{eq:9.5}) and (\ref{eq:9.6}) imply,
\begin{eqnarray}
\Phi_{o}^{(1)(z)}(\vec{x},\acute{\vec{x}})&=&-\frac{iqc\beta_{z}\eta}{\pi^{2}a^{2}}\sum_{m=-\infty}^{\infty}\sum_{n=1}^{\infty}x_{mn}\frac{J_{m}({x_{mn}\acute{\varrho}})}{J_{m+1}(x_{mn})}
\nonumber \\
&&\times\int_{-\infty}^{\infty}d\kappa e^{i\kappa(\zeta-\acute{\zeta})}e^{im(\phi-\acute{\phi})}\frac{sgn(\kappa)}{I_{m}(|\kappa|)} \nonumber \\
&&\times\frac{K_{m}(|\kappa|)I_{m}(|\kappa|\varrho)}{K_{m-1}(|\kappa|)+K_{m+1}(|\kappa|)}\frac{1}{x_{mn}^{2}+\kappa^{2}},\label{eq:9.6.1}
\end{eqnarray}
where, $\kappa=ka$, $\varrho=\frac{\rho}{a}$, $\acute{\varrho}=\frac{\acute{\rho}}{a}$, $\zeta=\frac{z}{a}$, and $\acute{\zeta}=\frac{\acute{z}}{a}$. Equation (\ref{eq:9.6.1}) is the first order electric potential inside the trap due to the motion of a particle moving along the cylinder axis. Note that '$(z)$' in the superscript indicates that the charge is moving along the $z$ axis (recall that $z$ axis is the cylinder axis).

\begin{figure}
\includegraphics{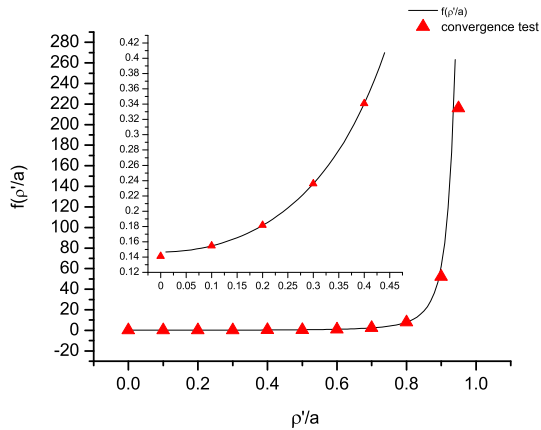}
\caption{\label{fig1} Numerical estimation for $f(\acute{\varrho})$ is presented. Inset is an expanded view near the origin to highlight the presence of the offset. $\rho=0$ is excluded due to numerical error. The settings are $max(m)=100$, $max(n)=5000$, $max(\kappa)=40$, and $d\kappa=0.01$. \textcolor{red}{$\blacktriangle$} represents $f(\acute{\varrho})$ evaluated with coarser settings for comparison. The coarse settings are $max(m)=50$, $max(n)=2500$, $max(\kappa)=20$, and $d\kappa=0.02$. This suggests that to a good approximation the convergence criteria is met at least for $\acute{\varrho}<0.9$. As it can be seen in the figure, beyond $0.9a$ there is numerical error due to the need for larger cutoffs. However, in real applications we are not concerned with these radius ranges.}
\end{figure}

Equations (\ref{eq:9.6.1}) and $E_{zo}^{(1)(z)}=-\frac{1}{a}\frac{\partial\Phi_{o}^{(1)(z)}}{\partial \zeta}$ give,
\begin{eqnarray}
E_{zo}^{(1)(z)}&=&-\frac{qc\beta_{z}\eta}{\pi^{2}a^{3}}\sum_{m=-\infty}^{\infty}\sum_{n=1}^{\infty}x_{mn}\frac{J_{m}({x_{mn}\acute{\varrho}})}{J_{m+1}(x_{mn})}
\nonumber \\
&&\times\int_{-\infty}^{\infty}d\kappa e^{i\kappa(\zeta-\acute{\zeta})}e^{im(\phi-\acute{\phi})}\frac{\kappa\cdot sgn(\kappa)}{I_{m}(|\kappa|)} \nonumber \\
&&\times\frac{K_{m}(|\kappa|)I_{m}(|\kappa|\varrho)}{K_{m-1}(|\kappa|)+K_{m+1}(|\kappa|)}\frac{1}{x_{mn}^{2}+\kappa^{2}},\label{eq:9.7}
\end{eqnarray}
This is the $z$ (subscript) component of the resistive electric field inside the trap due to the motion of the charge along the $z$ axis (superscript). The conduction cooling force is found using Eq. (\ref{eq:9.7}) evaluated at $\vec{x}=\acute{\vec{x}}$. Thus,
\begin{eqnarray}
F_{zo}^{(1)(z)}&=&-\frac{q^{2}c\beta_{z}\eta}{\pi^{2}a^{3}}\sum_{m=-\infty}^{\infty}\sum_{n=1}^{\infty}x_{mn}\frac{J_{m}({x_{mn}\acute{\varrho}})}{J_{m+1}(x_{mn})}
\nonumber \\
&&\times\int_{-\infty}^{\infty}d\kappa \frac{K_{m}(|\kappa|)I_{m}(|\kappa|\acute{\varrho})}{I_{m}(|\kappa|)} \nonumber \\
&&\times\frac{|\kappa|}{K_{m-1}(|\kappa|)+K_{m+1}(|\kappa|)}\frac{1}{x_{mn}^{2}+\kappa^{2}}\nonumber\\
&=&-\frac{q^{2}c\beta_{z}\eta}{\pi a^{3}}f(\acute{\varrho}).\label{eq:9.8}
\end{eqnarray}
The function $f(\acute{\varrho})$ is introduced for the sake of compactness and simplicity. A numerical estimation of this function is shown in Fig. (\ref{fig1}).

Considering the fact that Eq. (\ref{eq:8}) applies to all orders, it could be instructive to investigate how these perturbative terms combine together. We can rewrite Eq. (\ref{eq:8}) as
\begin{equation}
E_{\rho i}= c \beta_{z} \eta \frac{\partial}{\partial z}\left(\sigma^{(0)}+\sigma^{(1)}+\cdots\right), \label{eq:12}
\end{equation}
and by iterating the process we can find the higher order correction terms with respect to the zeroth order. Every higher order term extracts a factor of $\frac{\epsilon_{0}c\beta_{z}\eta}{a}$ which has to be small in order to make our perturbation approach legitimate. The condition, that
\begin{equation}
\frac{\epsilon_{0}c\beta_{z}\eta}{a}\ll 1,\label{eq:13}
\end{equation}
imposes a constraint on the velocity of the charge, resistivity and the radius of the trap.

Intuitively, we expect that if the charge moves close enough to the cylinder wall, it "sees" the wall as a flat surface. The first order resistive force due to a charge moving in parallel with a conductive flat surface is calculated by Boyer \cite{boyer1974penetration} in which he predicts that the the first order resistive force on the charge $q$ moving with the velocity of $v_{z}$ and the distance $d$ to the wall is
\begin{equation}
F^{(1)}_{plane}=-\frac{1}{16}\frac{q^{2}\eta}{\pi d^{3}}v_{z}.\label{eq:13.1}
\end{equation}
By dividing Eq. (\ref{eq:13.1}) by Eq. (\ref{eq:9.8}) we obtain,
\begin{eqnarray}
\frac{F_{z,wall}^{(1)}}{F_{z,cylinder}^{(1)}} &=& \frac{1}{16f(\acute{\varrho})}\left(\frac{a}{d}\right)^{3} \nonumber \\
&=&\frac{1}{16f(\acute{\varrho})}\left(\frac{1}{1-\acute{\varrho}}\right)^{3}.\label{eq:20_3}
\end{eqnarray}

\begin{figure}
\includegraphics{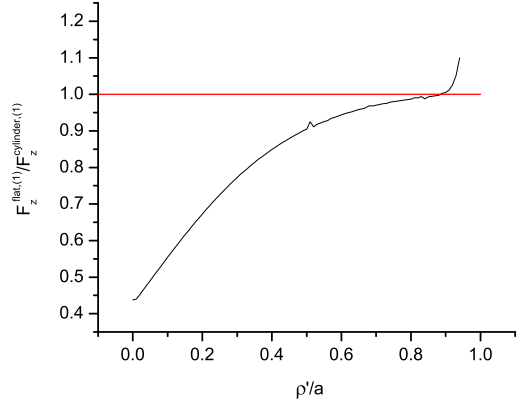}
\caption{\label{zRatio} $F_{z}^{flat,(1)}/F_{z}^{cylinder,(1)}$ shown as a function of charge distance to the wall. A trend towards unity is visible, however for $\frac{\acute{\rho}}{a}>0.9$ numerical instabilities become significant.}
\end{figure}

The ratio in Eq. (\ref{eq:20_3}) is presented in Fig. \ref{zRatio}. This shows that the ratio of the two quantities asymptotically approaches unity as the charge moves closer to the wall along the $z$ axis. Beyond the radius $0.9a$ the numerical errors become significant. Note that the closer the charge is to the cylinder wall, the more terms are needed in the summation over $m$. However in most practical purposes we are interested in the cases where the charges are further away from the wall.
\subsection{Charge motion along the cylinder radius}
In this section we will consider the motion of the charge along the radius of the cylinder.

In this case we can not exactly follow the first few steps similar to the $z$ axis motion because the surface charge density does not remain invariant under the translational motion and the geometry of the problem changes as the charge travels along the radius of the trap. Instead, using the chain rule, we can write
\begin{equation}
\frac{\partial\sigma}{\partial t}=c\beta_{\rho}\frac{\partial{\sigma}}{\partial{\acute{\rho}}},\label{eq:21}
\end{equation}
in which $\beta_{\rho}$ is $\frac{v_{\rho}}{c}$, and combining equations (\ref{eq:5}), (\ref{eq:6}), and (\ref{eq:21}) yields,
\begin{equation}
E_{\rho i}^{(1)}=-c\beta_{\rho}\eta\frac{\partial \sigma^{(0)}}{\partial\acute{\rho}}.\label{eq:22}
\end{equation}
This is the first order radial component of the velocity field at $\rho=a$ (just inside the conductor) which is in charge of bringing the right amount of charge to the surface of the wall as the charged particle travels along the radius.

Equations (\ref{eq:3}), (\ref{eq:9.2}), (\ref{eq:22}) and (\ref{eq:9:1}) (with $B_{m}=0$ for finite potential at large radius) imply,
\begin{eqnarray}
&&-\sum_{m=-\infty}^{\infty}\int_{-\infty}^{\infty}dk A_{m}(k)|k|\acute{K_{m}}(|k|\rho)e^{ikz}e^{im\phi} \nonumber \\
&&=-\frac{qc\beta_{\rho}\eta}{4\pi a^{3}}\sum_{m=-\infty}^{\infty}\sum_{n=1}^{\infty}x_{mn}\frac{J_{m-1}\left(\frac{x_{mn}\acute{\rho}}{a}\right)-J_{m+1}\left(\frac{x_{mn}\acute{\rho}}{a}\right)}
{J_{m+1}(x_{mn})} \nonumber \\
&& \times e^{\frac{-x_{mn}|z|}{a}}e^{im\phi}.\label{eq:22.1}
\end{eqnarray}
The same notation and conventions are employed here as used in Section (\ref{mo_axis}). Accordingly, the electrical potential expansion coefficients inside the conductor are
\begin{eqnarray}
A_{m}(k)&=&-\frac{qc\beta_{\rho}\eta}{2\pi^{2}a}\left(\frac{1}{|ka|(K_{m-1}(|ka|)+K_{m+1}(|ka|))}\right)\nonumber \\
&&\times\sum_{n=1}^{\infty}x^{2}_{mn}\frac{J_{m-1}\left(\frac{x_{mn}\acute{\rho}}{a}\right)-J_{m+1}\left(\frac{x_{mn}\acute{\rho}}{a}\right)}{J_{m+1}(x_{mn})
(x_{mn}^{2}+(ka)^{2})}.\label{eq:22.2}
\end{eqnarray}

The perturbative electric potential therefore can be expanded outside the conductor (inside the trap) similar to Eq. (\ref{eq:9.5}), and considering that the potentials on the boundary ($\rho=a$) need to be equal, we can repeat the statement in Eq. (\ref{eq:9.6}).

Therefore the first order perturbative electric potential, inside the trap, due to the charge motion along the cylinder radius is
\begin{eqnarray}
\Phi_{o}^{(1)(\rho)}(\vec{x},\acute{\vec{x}})&=&-\frac{qc\beta_{\rho}\eta}{2\pi^{2}a^{2}}\sum_{m=-\infty}^{\infty}\sum_{n=1}^{\infty}x_{mn}^{2}
\nonumber \\
&&\times\frac{J_{m-1}\left(x_{mn}\acute{\varrho}\right)-J_{m+1}\left(x_{mn}\acute{\varrho}\right)}{J_{m+1}(x_{mn})}\nonumber \\
&&\times\int_{-\infty}^{\infty}d\kappa\frac{e^{i\kappa(\zeta-\acute{\zeta})}e^{im(\phi-\acute{\phi})}}{K_{m-1}(|\kappa|)+K_{m+1}(|\kappa|)} \nonumber \\
&&\times\frac{K_{m}(|\kappa|)}{I_{m}(|\kappa|)}\frac{I_{m}(|\kappa|\varrho)}
{|\kappa|(x_{mn}^{2}+\kappa^{2})}.\label{eq:22.2.1}
\end{eqnarray}

Equation (\ref{eq:22.2.1}) and $I_{m}(x)+I_{m}(x)=2\acute{I}_{m}(x)$ yields,
\begin{eqnarray}
E_{\rho o}^{(1)(\rho)}&=&\frac{qc\beta_{\rho}\eta}{4\pi^{2}a^{3}}\sum_{m=-\infty}^{\infty}\sum_{n=1}^{\infty}x_{mn}^{2}
\nonumber \\
&&\times\frac{J_{m-1}\left(x_{mn}\acute{\varrho}\right)-J_{m+1}\left(x_{mn}\acute{\varrho}\right)}{J_{m+1}(x_{mn})}\nonumber \\
&&\times\int_{-\infty}^{\infty}d\kappa \frac{e^{i\kappa(\zeta-\acute{\zeta})}e^{im(\phi-\acute{\phi})}}{K_{m-1}(|\kappa|)+K_{m+1}(|\kappa|)} \nonumber \\
&&\times\frac{K_{m}(|\kappa|)}{I_{m}(|\kappa|)}\frac{I_{m-1}(|\kappa|\varrho)+I_{m-1}(|\kappa|\varrho)}
{(x_{mn}^{2}+\kappa^{2})},\label{eq:22.3}
\end{eqnarray}
which is the $\rho$ (subscript) component of the resistive electric field outside the conductor (inside the trap) due to the motion of the charge along the $\rho$ axis (superscript).

The resistive force can be found using Eq. (\ref{eq:22.3}) evaluated at $\vec{x}=\acute{\vec{x}}$. Therefore, the resistive force is
\begin{eqnarray}
F_{\rho o}^{(1)(\rho)}&=&qE_{\rho o}^{(1)}|_{\vec{x}=\acute{\vec{x}}}\nonumber\\
&=&-\frac{q^{2}c\beta_{\rho}\eta}{\pi a^{3}}g(\acute{\varrho}),\label{22.4}
\end{eqnarray}
where the function $g$ is used, similar to the previous section, for the sake of brevity and simplicity.

\begin{figure}
\includegraphics{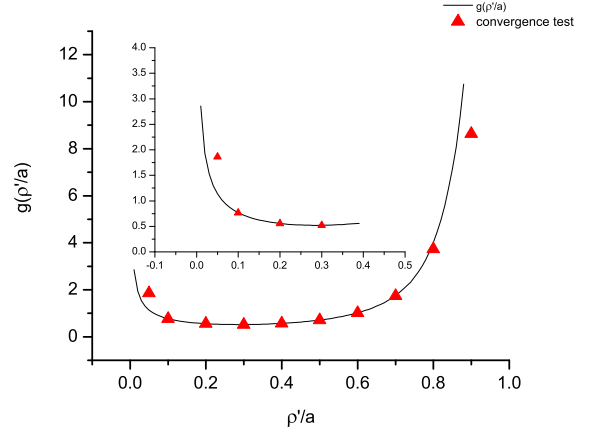}
\caption{\label{fig2}Numerical estimation of $g(\acute{\varrho})$ is presented. The inset is a larger view of the function. The function is not evaluated at $\rho=0$ and beyond $\rho=0.9$ due to numerical error. The settings in this case are $max(m)=100$, $max(n)=5000$, $max(\kappa)=30$, and $d\kappa=0.01$. \textcolor{red}{$\blacktriangle$} represents $g(\acute{\varrho})$ evaluated using coarser settings for comparison. In this case the settings are $max(m)=50$, $max(n)=2500$, $max(\kappa)=20$, and $d\kappa=0.02$. This suggests that to a good approximation the convergence criteria is well met for $0.1<\acute{\varrho}<0.8$.}
\end{figure}

A numerical approximation of $g(\acute{\varrho})$ is presented in Fig. (\ref{fig2}). The increase in the function near the origin can possibly be explained by the formation of a dipole along(opposite-to) the motion of a positively(negatively) charged particle while the particle senses more both the repulsion from the positive(negative) pole in the front and attraction from the negative(positive) one behind.
\subsection{Charge motion along the cylinder azimuthal angle}
Just like the two cases above, any motion along $\hat{\phi}$ will also create a resistive electric field which causes a damping force opposing the motion of the charge. Similar to what we have done for the $z$ axis motion, we can write
\begin{equation}
\frac{\partial \sigma}{\partial t}=-\dot{\phi}\frac{\partial\sigma}{\partial\phi}.\label{eq:25}
\end{equation}
Using equations (\ref{eq:5}), (\ref{eq:6}), and (\ref{eq:25}), again we solve for the first order perpendicular electric field on the boundary and just inside the conductor,
\begin{equation}
E_{\rho i}^{(1)}=\eta\dot{\phi}\frac{\partial\sigma^{(0)}}{\partial\phi}.\label{eq:26}
\end{equation}
Equations (\ref{eq:3}), (\ref{eq:9.5}) and (\ref{eq:26}) yield
\begin{eqnarray}
&&-\sum_{m=-\infty}^{\infty}\int_{-\infty}^{\infty}dk A_{m}(k)|k|\acute{K_{m}}(|k|\rho)e^{ikz}e^{im\phi}\nonumber\\
&&=-\frac{iq\eta}{2\pi a^{2}}\dot{\phi}
\sum_{m=-\infty}^{\infty}\sum_{n=1}^{\infty}
m e^{im\phi} e^{\frac{-x_{mn}|z|}{a}}
\frac{J_{m}\left(\frac{x_{mn}\acute{\rho}}{a}\right)}{J_{m+1}(x_{mn})},\label{eq:27}
\end{eqnarray}
and as a result, the coefficients can be written as
\begin{eqnarray}
A_{m}(k)&=&-\frac{iq\eta}{\pi^{2}}\dot{\phi}\left(\frac{1}{|ka|(K_{m-1}(|ka|)+K_{m+1}(|ka|))}\right)\nonumber \\
&&\times\sum_{n=1}^{\infty}mx_{mn}\frac{J_{m}\left(\frac{x_{mn}\acute{\rho}}{a}\right)}{J_{m+1}(x_{mn})
(x_{mn}^{2}+(ka)^{2})}.\label{eq:27.1}
\end{eqnarray}

The perturbative electric potential inside the trap can be expanded similar to Eq. (\ref{eq:9.5}) while only keeping $B_{m}$ nonzero. Equation (\ref{eq:27.1}) and the equality of the potentials on the boundary using Eq. (\ref{eq:9.6}) imply
\begin{eqnarray}
\Phi_{\phi o}^{(1)(\phi)}(\vec{x},\acute{\vec{x}})&=&-\frac{iqc\beta_{\phi}\eta}{\pi^{2}\acute{\varrho}a^{2}}\sum_{m=-\infty}^{\infty}\sum_{n=1}^{\infty}m x_{mn}
\frac{J_{m}\left(x_{mn}\acute{\varrho}\right)}{J_{m+1}(x_{mn})}\nonumber \\
&&\times\int_{-\infty}^{\infty}d\kappa
\frac{e^{i\kappa(\zeta-\acute{\zeta})}e^{im(\phi-\acute{\phi})}}{K_{m-1}(|\kappa|)+K_{m+1}(|\kappa|)}\nonumber \\
&&\times\frac{K_{m}(|\kappa|)}{I_{m}(|\kappa|)}\frac{I_{m}(|\kappa|\varrho)}
{(x_{mn}^{2}+\kappa^{2})},\label{eq:27.2.1}
\end{eqnarray}

\begin{figure}
\includegraphics{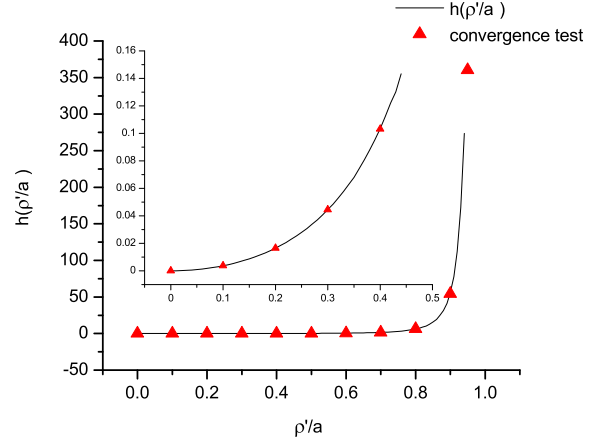}
\caption{\label{fig3}The function, $h(\acute{\varrho})$ is evaluated numerically and presented. The inset is a larger view of the function. The settings for this evaluation are $max(m)=100$, $max(n)=5000$, $max(\kappa)=40$, and $d\kappa=0.01$. \textcolor{red}{$\blacktriangle$} represents $h(\acute{\varrho})$ evaluated using coarser settings for comparison. The settings are $max(m)=50$, $max(n)=2500$, $max(\kappa)=20$, and $d\kappa=0.02$. This suggests that to a good approximation the convergence criteria is met for $\acute{\varrho}<0.9$. However, at $\acute{\varrho}=0.95$ there is significant numerical error which is due to need for higher cutoffs in the summations. However, for practical purposes we are not concerned with those ranges.}
\end{figure}

Thus the $\phi$ component of the resistive electric field due to a charge moving along $\hat{\phi}$ is
\begin{eqnarray}
E_{\phi o}^{(1)}&=&-\frac{qc\beta_{\phi}\eta}{\pi^{2}\acute{\varrho}^{2}a^{3}}\sum_{m=-\infty}^{\infty}\sum_{n=1}^{\infty}m^{2}x_{mn}
\frac{J_{m}\left(x_{mn}\acute{\varrho}\right)}{J_{m+1}(x_{mn})}\nonumber \\
&&\times\int_{-\infty}^{\infty}d\kappa e^{i\kappa(\zeta-\acute{\zeta})}e^{im(\phi-\acute{\phi})}
\frac{1}{K_{m-1}(|\kappa|)+K_{m+1}(|\kappa|)}\nonumber \\
&&\times\frac{K_{m}(|\kappa|)}{I_{m}(|\kappa|)}\frac{I_{m}(|\kappa|\varrho)}
{(x_{mn}^{2}+\kappa^{2})},\label{eq:27.2}
\end{eqnarray}
The resistive force is calculated using Eq. (\ref{eq:27.2}) evaluated at $\vec{x}=\acute{\vec{x}}$. Therefore, the resistive force is
\begin{eqnarray}
F_{\phi o}^{(1)}&=&qE_{\phi o}^{(1)}|_{\vec{x}=\acute{\vec{x}}}\nonumber\\
&=&-\frac{q^{2}c\beta_{\phi}\eta}{\pi\acute{\varrho}^{2}a^{3}}h(\acute{\varrho}),\label{27.3}
\end{eqnarray}
where $h$ is numerically estimated and shown in Fig. (\ref{fig3}).

Similar to the axial motion case, we expect that when the charge moves along the $\hat{\phi}$ and in parallel and close to the wall, the resistive force asymptotically approaches the flat surface limit. Thus we expect that
\begin{eqnarray}
\lim_{\acute{\varrho}\rightarrow 1} \frac{F_{z,wall}^{(1)}}{F_{z,cylinder}^{(1)}} &=& \lim_{\acute{\varrho}\rightarrow 1}\frac{\acute{\varrho}^{2}}{16h(\acute{\varrho})}\left(\frac{a}{d}\right)^{3} \nonumber \\
&=&\lim_{\acute{\varrho}\rightarrow 1}\frac{1}{16h(\acute{\varrho})}\left(\frac{1}{1-\acute{\varrho}}\right)^{3}.\label{eq:27_4}
\end{eqnarray}

\begin{figure}
\includegraphics{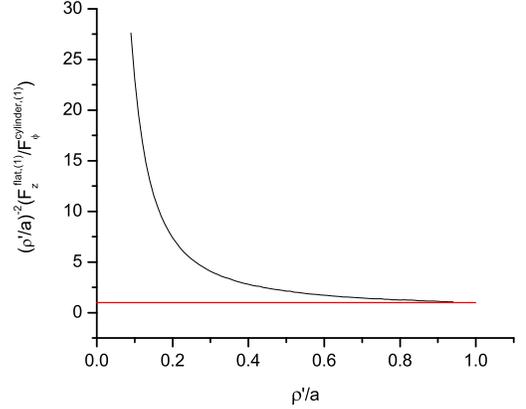}
\caption{\label{phiRatio} $\acute{\varrho}^{-2} F_{z}^{flat,(1)}/F_{\phi}^{cylinder,(1)}$ shown as a function of charge distance to the wall. The horizontal line represents $y=1$. A trend towards unity is visible. $\acute{\varrho}^{-2}$ is factored out for faster numerical convergence and less numerical instability.}
\end{figure}

The ratio $\frac{1}{16h(\acute{\varrho})}\left(\frac{1}{1-\acute{\varrho}}\right)^{3}$ as a function of $\acute{\varrho}$ is presented in Fig. \ref{phiRatio}. As clearly shown in this figure, the ratio indeed approaches unity suggesting that the $\phi$ component of resistive force on the charge approaches the flat surface limit as it moves closer to the cylinder wall. The reason $\acute{\varrho}^{2}$ is factored out is to avoid needing more terms in the series to show the convergence, however we know that since $\acute{\varrho}^{2}$ approaches one as $\acute{\varrho}$ approaches unity and it does not affect our conclusion.
\subsection{Compact form}
Now that we have found all the proper single particle conduction cooling forces due to different degrees of freedom inside the cylinder, we can use them to define a tensor which gives us the resistive force as we multiply it by the particle velocity vector as
\begin{equation}
f_{\alpha}=-\eta_{\alpha\beta}(\acute{\varrho})v^{\beta},\label{eq:30}
\end{equation}
where $f_{\alpha}$ is the cooling force, $\eta_{\alpha\beta}$ is the conduction cooling tensor which due to the cylindrical symmetry is a function of the radial position of the charge, and $v^{\beta}$ which is the velocity of the charge. $\eta_{\alpha\beta}$ is a diagonal tensor since we have already picked its eigen-vectors as our coordinate system. The non-zero components of the tensor are
\begin{subequations}
\label{compact}
\begin{eqnarray}
\eta_{\rho\rho}=
\frac{q^{2}\eta}{\pi a^{3}}g(\acute{\varrho}),
\label{eq:31a}
\\
\eta_{\phi\phi}=
\frac{q^{2}\eta}{\pi\acute{\varrho}^{2}a^{3}} h(\acute{\varrho}),
\label{eq:31b}
\\
\eta_{zz}=
\frac{q^{2}\eta}{\pi a^{3}}f(\acute{\varrho}).
\label{eq:31c}
\end{eqnarray}
\end{subequations}
Therefore,
\begin{eqnarray}
\eta_{\alpha\beta}&=&
\frac{q^{2}\eta}{\pi a^{3}}\biggl(g(\acute{\varrho})\delta_{\alpha\rho}+
\frac{1}{\acute{\varrho}^{2}}h(\acute{\varrho})\delta_{\alpha\phi}+ \nonumber \\
&&f(\acute{\varrho})\delta_{\alpha z}\biggr)\delta_{\alpha\beta},
\label{eq:32}
\end{eqnarray}
in which $\delta_{\alpha\beta}$ is the Kronecker delta.

\section{\label{sec:iii}Method and Calculation: Many Particle Analysis}
In the previous section, we have calculated the cooling force on a moving charge inside a long conductive cylinder. We can use our results to formulate the effect for a many particle system. Linearity of the Maxwell's equations allows us to sum on the perturbative electric potentials and the electric fields created by different moving charges.

So far, we have found the resistive electric potential and the corresponding electric field generated by a moving charge, and evaluated the electric field at the position of the charge to calculate the self-cooling force. However, the resistive electric field not only affects the source particle but also it applies some force on other particles as well. Therefore, to find the total cooling effect, we need to sum over all of the interactions in the ensemble.

For convenience in our formulation, we define three potentials corresponding to equations (\ref{eq:9.6.1}), (\ref{eq:22.2.1}) and (\ref{eq:27.2.1}) which are normalized with the particle velocity ($c\beta_{\alpha}$), the charge $q$ and the conductivity of the cylinder $\eta$. The reason is to make these potentials only geometry dependent so that we can easily scale the final results. Thus,
\begin{subequations}
\label{compact} 
\begin{eqnarray}
\varphi^{\rho} &=&\frac{1}{qc\eta}\frac{\Phi_{o}^{(1)(\rho)}}{\beta_{z}},\label{eq:32.1.a}
\\
\varphi^{\phi} &=&\frac{1}{qc\eta}\frac{\Phi_{o}^{(1)(\phi)}}{\beta_{\phi}},\label{eq:32.1.b}
\\
\varphi^{z} &=&\frac{1}{qc\eta}\frac{\Phi_{o}^{(1)(z)}}{\beta_{z}}.\label{eq:32.1.c}
\end{eqnarray}
\end{subequations}
The subscript $o$ is dropped since we are only interested in the potentials and the fields outside the conductor (inside the trap). The superscript "$(1)$" is also dropped since the first order approximation is consistently used. We define the tensor
\begin{equation}
\varepsilon_{\alpha\beta}(\vec{x},\acute{\vec{x}})=-\left[
\begin{array}{rrr} \varphi^{\rho}_{,\rho} & \varphi^{\phi}_{,\rho} & \varphi^{z}_{,\rho} \\
                   \frac{1}{\rho}\varphi^{\rho}_{,\phi} & \frac{1}{\rho}\varphi^{\phi}_{,\phi} & \frac{1}{\rho}\varphi^{z}_{,\phi} \\
                   \varphi^{\rho}_{,z} & \varphi^{\phi}_{,z} & \varphi^{z}_{,z}
\end{array}
\right]\label{eq:32.2}
\end{equation}
in which the derivative convention is being used $\varphi_{,x}=\frac{\partial \varphi}{\partial x}$.

Therefore, the resistive electric field evaluated at $\vec{x}$ due to a moving charge $q$ positioned at $\acute{\vec{x}}$ with the velocity $\acute{v}^{\beta}$ is
\begin{equation}
E_{\alpha}(\vec{x})=q\eta\varepsilon_{\alpha\beta}(\vec{x},\acute{\vec{x}})\acute{v}^{\beta}.\label{eq:32.3}
\end{equation}

As a result, the power applied to a second charge $q$ at $\vec{x}$ with the velocity $v^{\alpha}$ due to this electric field is
\begin{equation}
p=q^{2}\eta\varepsilon_{\alpha\beta}(\vec{x},\acute{\vec{x}})v^{\alpha}\acute{v}^{\beta}.\label{eq:32.4}
\end{equation}
Note that this power could be positive or negative, however the resistive electric field is maximum at the position of the source charge such that overall the resistive effect removes energy from the ensemble. The total cooling power applied to the ensemble is obtained by summing over all the mutual interactions between the particles. Using the ensemble distribution function $f(\vec{x},\vec{v};t)$ which is the density of the particles at the position $\vec{x}$ and with the velocity $\vec{v}$ at time $t$, the total dissipative power $P$ is
\begin{eqnarray}
P(t)&=&q^{2}\eta\int d^{3}rd^{3}v\int d^{3}\acute{r}d^{3}\acute{v} \nonumber \\
&&\times f(\vec{x},\vec{v};t)f(\acute{\vec{x}},\acute{\vec{v}};t)\varepsilon_{\alpha\beta}(\vec{x},\acute{\vec{x}})v^{\alpha}\acute{v}^{\beta}.\label{eq:32.5}
\end{eqnarray}
The time evolution of the ensemble leads to $f(\vec{x},\vec{v};t)$, and correspondingly, the distribution gives $P(t)$. The dynamics of the ensemble can be found using a plasma simulation code such as WARP[\textcolor{red}{citation needed}]. However, such codes do not take into account the image charge cooling effects. In principle, these interactions can be quantified in each time-step during the simulation.

Note that in order to derive Eq. (\ref{eq:32.5}), our fundamental assumption is the linearity of Maxwell's equations and therefore, we can add the fields due to individual particles to obtain the collective field. Therefore, Eq. (\ref{eq:32.5}) can be used for an ensemble with an arbitrary density.
\subsection{\label{sec:i}cooling a weakly interacting ensemble}
If we have an ensemble such that the resistive field produced by each particle is small enough at the position of the other particles, we can reduce the many particle problem to a single particle problem. This can be useful for example in a pre-trap section in which the particles cool down to the favorable energy before being injected into the trap. For more simplification, we only consider the axial motion cooling in which we only use $\varepsilon_{zz}$.

\begin{figure}
\includegraphics{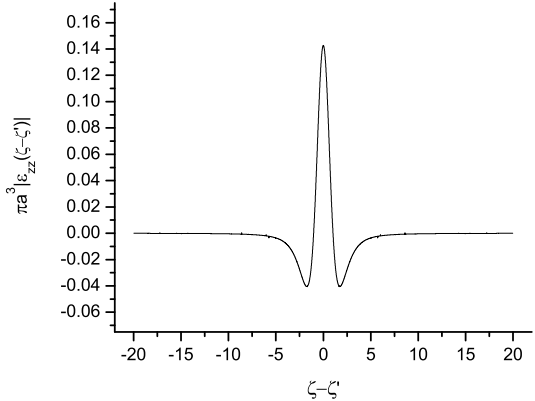}
\caption{\label{ePsil} Estimation $\varepsilon_{zz}$ evaluated on the $z$ axis.}
\end{figure}

A numerical estimation of $\varepsilon_{zz}$ on axis is being presented in Fig. \ref{ePsil}. It can clearly be seen in the figure that $\varepsilon_{zz}$ is fairly small beyond the range of $10a$ around $\zeta-\acute{\zeta}$. Therefore, if the average distance between the particles is $10a$ or more, the mutual interactions between the particles will be relatively small. Thus each particle will cool down approximately similar to a single particle inside the pre-trap section.

If we optimize the resistivity of the pre-trap, we can cool down the particles enough in a reasonably short period time before injecting them into the trap. We will look at the calculated cooling time constants in section \ref{results} and compare them to radiation and gas cooling which are being commonly used for positron cooling.

In the trap of our interest, the radius of each micro-tube is $50 \mu m$. Thus, tens of particles per micro-tube can be cooled down per cooling cycle in a $5 cm$ long pre-trap section. Recall that in a micro-trap, tens thousands of these tubes are packed together. Therefore, using a pre-trap with the same number of micro-tubes aligned with the trap, hundreds of thousands of particles can be injected into the trap in each cycle.
\section{\label{results}Results}
In the following paragraphs we will compare the cooling time using major positron cooling methods including gas cooling and radiation cooling with conduction cooling.

Gas cooling has been employed extensively and, more specifically, it has been used for positron cooling by Surko \textit{et al.} and others \cite{surko2004emerging,clarke2006design} where N$_2$ is typically employed. Although other gases have been used with less pressures for other purposes, we have calculated the thermalization time for them at the commonly used pressures for the sake of comparison. In the experiments reported by Al-Qaradawi \textit{et al.}, the thermalization time of sub-$eV$ positrons for different molecular gasses has been measured \cite{Ilham2000thermalization} and used to estimate the thermalization time for different gasses at room temperature and the pressures (\num{1.3e-3}mbar\cite{surko2004emerging}), and ($\num{1.5E-5}$ mbar\cite{clarke2006design}). In Table \ref{table:1} the thermalization time for sub-eV positrons in different gases including N$_2$ has been enlisted for comparison. However, annihilation with the electrons of the gas molecules decreases the efficiency of the process. This depends on the energy of the positrons, the state, and the composition of the gas. For example gas cooling efficiency has been reported $15\%-26\%$ using $N_{2}$ depending on the pressure \cite{PhysRevA.46.5696}.
\begin{table}
\centering
\caption{\label{table:1}Gas cooling thermalization time $\tau_{GC}$ for sub-$eV$ positrons to room temperature for different gasses at room temperature.}
\begin{ruledtabular}
\begin{tabular}{l*{5}{c}r}
&\multicolumn{4}{c}{Pressures(mbar)} \\
\cline{2-5}
Gas               && \num{1.3E-3} && \num{1.5E-5} \\
\hline 
H$_2$                && \num{1.66}$(ms)$  && \num{121.5}$(ms)$  \\
CH$_4$               && \num{0.207}$(ms)$  && \num{15.1}$(ms)$  \\
CO                && \num{0.761}$(ms)$ && \num{55.7}$(ms)$ \\
CO$_2$               && \num{69.1}$(\mu s)$ && \num{5.06}$(ms)$ \\
N$_2$O               && \num{69.1}$(\mu s)$ && \num{5.06}$(ms)$ \\
SF$_6$               && \num{27.6}$(\mu s)$ && \num{2.02}$(ms)$ \\
N$_2$                && \num{9.68}$(ms)$ && \num{709}$(ms)$ \\
\end{tabular}
\end{ruledtabular}
\end{table}

Although radiation cooling may not be applicable for micro-traps, it might be useful to compare its cooling time constant with other methods. The equation for the temperature can be written as \cite{beck:1250}
\begin{equation}
\frac{dT}{dt}=-\frac{T}{\tau_{RC}}, \label{eq:39}
\end{equation}
where
\begin{equation}
\tau_{RC}=\frac{9\pi m\epsilon_{0}c^{3}}{2e^{2}\Omega_{c}^{2}} \label{eq:39.1}
\end{equation}
in SI units. Equation (\ref{eq:39}) holds as long as the heat bath temperature is much less than the ensemble temperature and quantum effects are negligible \cite{BentPhd05}. For convenience $\tau_{RC}$ is tabulated in Table \ref{table:2} for different magnetic fields.
\begin{table}
\centering
\caption{\label{table:2}Radiation cooling thermalization time $\tau_{RC}$ for 0.5eV positrons to room temperature in different magnetic fields.}
\begin{ruledtabular}
\begin{tabular}{l*{5}{c}r}
&\multicolumn{4}{c}{Magnetic fields} \\
\cline{2-5}
         & 10mT & 1T & 3T & 7T \\
\hline 
$\tau_{RC}$(s)              & \num{3.8e4} & \num{3.8} & \num{0.429} & \num{7.9e-2} \\
\end{tabular}
\end{ruledtabular}
\end{table}

As shown in Section \ref{sec:i}, conduction cooling of an ensemble with a low enough density in a pre-trap section can be approximated using single particle cooling. For simplicity, we only consider the axial motion cooling. The cooling force on a particle moving along the trap axis can be found using Eq. (\ref{eq:9.8}). This causes an exponentially damped motion with the energy loss time constant of
\begin{equation}
\tau_{CC}=\frac{m\pi a^{3}}{2q^{2}\eta}\frac{1}{f(\acute{\varrho})}, \label{eq:39.2}
\end{equation}
for a low density ensemble consisting of particles with charge $q$ and mass $m$ inside a conductive micro-tube of radius $a$ and the resistivity $\eta$. $f(\acute{\varrho})$ is presented in Fig. \ref{fig1} in which $\acute{\varrho}$ is the radial position of the charge ($\acute{\rho}$) divided by the radius of the trap ($a$).

The cooling time constants are presented for electrons/positrons for different pre-trap resistivities and radii in Table \ref{table:3}.
\begin{table}
\centering
\caption{\label{table:3}Conduction cooling time constant $\tau_{CC}$ for electrons/positrons for different conductors and trap radiuses.}
\begin{ruledtabular}
\begin{tabular}{l*{5}{c}r}
&\multicolumn{3}{c}{Trap radius} \\
\cline{2-4}
Conductor         & 50($\mu m$) & 10($\mu m$) & 1($\mu m$) \\
\hline 
Carbon\footnote[2]{Amorphous Carbon}            & \num{81}$(ms)$  & \num{0.65}$(ms)$  & \num{0.65}$(\mu s)$ \\
GaAs             & \num{4.87}$(ms)$  & \num{39}$(\mu s)$  & \num{39}$(ns)$ \\
Germanium        & \num{106} $(\mu s)$  & \num{0.85} $(\mu s)$ & - \\
\end{tabular}
\end{ruledtabular}
\end{table}

\section{discussion}
\subsection{Restatement of assumptions}
In this paper we have calculated the first order conduction cooling force, and its corresponding tensor inside a long aspect ratio conductive cylinder (trap) with a thick enough wall. These velocity fields drop to less than their $10\%$ value in a fraction of the trap radius, and therefore in practical purposes the trap walls are thick enough in our approximation. However, considering a small thickness for the trap wall is more mathematically involved and can be investigated in the future work. The finite thickness problem has been solved for a flat surface \cite{boyer1996penetration}, and accordingly, the expression for the conduction cooling force for a charge moving in parallel with a planar slab of thickness $l$ is
\begin{equation}
F_{slab}^{(1)}=-\frac{\eta e^{2} c\beta}{2\pi}
\left[\frac{1}{[2d]^{3}}+2\sum_{n=1}^{\infty}
\frac{1}{[2d+2nl]^{3}}\right], \label{eq:42}
\end{equation}
Surprisingly, the result show an increase in the resistive force as the thickness decreases although the first perturbation criteria puts a lower bound on $l$. Based on this fact, we would expect to see an increase in the resistive forces as the cylinder wall becomes thinner, although it needs to be proved and quantified.

Our final results in the forces and the cooling rates is independent of the relative permittivity of the conductors, although, in Eq. (\ref{eq:13}) care must be taken that such that the condition holds for larger permittivities. However, in case of metals the 'true' relative permittivity which is due to the dipolar interactions of the bound electrons, is close to one at low frequencies \cite{lourtioz2008photonic}.

The trap walls are assumed to be perfect with no surface roughness and no patch effect present. Considering the surface roughness including the wafer misalignments (a few microns to a few tens of microns in size) and the roughnesses because of the etching process (a fraction of micron to a few microns in size depending on the etching quality) will have several effects on the trap including changing the waveguide cut frequency, applying localized torque on the plasma and causing anomalies and instabilities, and changes in the conduction cooling rate. Developments based on any of these considerations requires more derivations and simulations which can be motivations for future work.
\subsection{Conclusions}
The first order field corrections due to the motion of a single point charge inside a long hollow conductor cylindrical trap is derived and numerically estimated. The first order single particle cooling force due to this motion resistive electric field has a tensorial relation ship of $f_{\alpha}=-\eta_{\alpha\beta}(\acute{\varrho})v^{\beta}$ where $\eta_{\alpha\beta}$ is the conduction cooling tensor and is proportional to $q^{2}\eta/a^{3}$, in which $q$ is the particle charge, $\eta$ is the trap electrical resistivity, and $a$ is the trap radius. This indicates that the conduction cooling is increasingly more effective in smaller trap radii.

In the case of axial and azimuthal motion, if the charge moves very close to the surface of the conductor, the cooling force should approach the force on a charge moving in parallel with a flat conductor and our results show that this is indeed the case.

Linearity of the electromagnetic theorem implies that the collective behaviour of particles can be obtained simply be summing over the velocity fields caused by different charges and motions. Therefore, we formulated the dissipation power for a collection of charges using $\varepsilon_{\alpha\beta}(\vec{x},\acute{\vec{x}})$ which is a geometry dependent tensor field. Equation (\ref{eq:32.5}) can be evaluated in association with a plasma dynamics simulation code such as WARP code \textcolor{red}{citation needed} since it is impossible to be done analytically unless the distribution function $f(\acute{\vec{x}},\acute{\vec{v}};t)$ is known.

Our low density approximation shows that the low density ensembles cool down with the single particle cooling rate. Therefore in a pre-trap with tens of thousands micro-tubes hundreds of thousands of particles can be cooled down in each cooling cycle. This is assuming only tens of particles are being cooled down in each micro-tube. For example in the $50 \mu m$ in radius micro-tubes with the electrical resistivity of $0.46 \Omega m$ the cooling time constant is $106 \mu s$.
\begin{acknowledgments}
We would like to thank program managers Dr. William Beck and Dr. Parvez Uppal of the Army Research Laboratory who provide funding under contract W9113M-09-C-0075, Positron Storage for Space and Missile Defense Applications, and program manager Dr. Scott Coombe of the Office of Naval Research who provide finding under award \#N00014-10-1-0543, Micro- and Nano-Traps to Store Large Numbers of Positron Particles at Very Large Densities. We also wish to thank M. Alizadeh for his useful discussions on the optimization of the programs.
\end{acknowledgments}

\bibliography{myreferences}

\begin{thebibliography}{34}%
\makeatletter
\providecommand \@ifxundefined [1]{%
 \@ifx{#1\undefined}
}%
\providecommand \@ifnum [1]{%
 \ifnum #1\expandafter \@firstoftwo
 \else \expandafter \@secondoftwo
 \fi
}%
\providecommand \@ifx [1]{%
 \ifx #1\expandafter \@firstoftwo
 \else \expandafter \@secondoftwo
 \fi
}%
\providecommand \natexlab [1]{#1}%
\providecommand \enquote  [1]{``#1''}%
\providecommand \bibnamefont  [1]{#1}%
\providecommand \bibfnamefont [1]{#1}%
\providecommand \citenamefont [1]{#1}%
\providecommand \href@noop [0]{\@secondoftwo}%
\providecommand \href [0]{\begingroup \@sanitize@url \@href}%
\providecommand \@href[1]{\@@startlink{#1}\@@href}%
\providecommand \@@href[1]{\endgroup#1\@@endlink}%
\providecommand \@sanitize@url [0]{\catcode `\\12\catcode `\$12\catcode
  `\&12\catcode `\#12\catcode `\^12\catcode `\_12\catcode `\%12\relax}%
\providecommand \@@startlink[1]{}%
\providecommand \@@endlink[0]{}%
\providecommand \url  [0]{\begingroup\@sanitize@url \@url }%
\providecommand \@url [1]{\endgroup\@href {#1}{\urlprefix }}%
\providecommand \urlprefix  [0]{URL }%
\providecommand \Eprint [0]{\href }%
\providecommand \doibase [0]{http://dx.doi.org/}%
\providecommand \selectlanguage [0]{\@gobble}%
\providecommand \bibinfo  [0]{\@secondoftwo}%
\providecommand \bibfield  [0]{\@secondoftwo}%
\providecommand \translation [1]{[#1]}%
\providecommand \BibitemOpen [0]{}%
\providecommand \bibitemStop [0]{}%
\providecommand \bibitemNoStop [0]{.\EOS\space}%
\providecommand \EOS [0]{\spacefactor3000\relax}%
\providecommand \BibitemShut  [1]{\csname bibitem#1\endcsname}%
\let\auto@bib@innerbib\@empty
\bibitem [{\citenamefont {Penning}(1936)}]{Penning1936873}%
  \BibitemOpen
  \bibfield  {author} {\bibinfo {author} {\bibfnamefont {F.}~\bibnamefont
  {Penning}},\ }\href {\doibase 10.1016/S0031-8914(36)80313-9} {\bibfield
  {journal} {\bibinfo  {journal} {Physica}\ }\textbf {\bibinfo {volume} {3}},\
  \bibinfo {pages} {873 } (\bibinfo {year} {1936})}\BibitemShut {NoStop}%
\bibitem [{\citenamefont {Malmberg}\ and\ \citenamefont
  {deGrassie}(1975)}]{malmberg1975properties}%
  \BibitemOpen
  \bibfield  {author} {\bibinfo {author} {\bibfnamefont {J.~H.}\ \bibnamefont
  {Malmberg}}\ and\ \bibinfo {author} {\bibfnamefont {J.~S.}\ \bibnamefont
  {deGrassie}},\ }\href {\doibase 10.1103/PhysRevLett.35.577} {\bibfield
  {journal} {\bibinfo  {journal} {Phys. Rev. Lett.}\ }\textbf {\bibinfo
  {volume} {35}},\ \bibinfo {pages} {577} (\bibinfo {year} {1975})}\BibitemShut
  {NoStop}%
\bibitem [{\citenamefont {Itano}\ \emph {et~al.}(1995)\citenamefont {Itano},
  \citenamefont {Bergquist}, \citenamefont {Bollinger},\ and\ \citenamefont
  {Wineland}}]{wineland_cooling1995}%
  \BibitemOpen
  \bibfield  {author} {\bibinfo {author} {\bibfnamefont {W.~M.}\ \bibnamefont
  {Itano}}, \bibinfo {author} {\bibfnamefont {J.~C.}\ \bibnamefont
  {Bergquist}}, \bibinfo {author} {\bibfnamefont {J.~J.}\ \bibnamefont
  {Bollinger}}, \ and\ \bibinfo {author} {\bibfnamefont {D.~J.}\ \bibnamefont
  {Wineland}},\ }\href {http://stacks.iop.org/1402-4896/1995/i=T59/a=013}
  {\bibfield  {journal} {\bibinfo  {journal} {Physica Scripta}\ }\textbf
  {\bibinfo {volume} {1995}},\ \bibinfo {pages} {106} (\bibinfo {year}
  {1995})}\BibitemShut {NoStop}%
\bibitem [{\citenamefont {Larson}\ \emph {et~al.}(1986)\citenamefont {Larson},
  \citenamefont {Bergquist}, \citenamefont {Bollinger}, \citenamefont {Itano},\
  and\ \citenamefont {Wineland}}]{PhysRevLett.57.70}%
  \BibitemOpen
  \bibfield  {author} {\bibinfo {author} {\bibfnamefont {D.~J.}\ \bibnamefont
  {Larson}}, \bibinfo {author} {\bibfnamefont {J.~C.}\ \bibnamefont
  {Bergquist}}, \bibinfo {author} {\bibfnamefont {J.~J.}\ \bibnamefont
  {Bollinger}}, \bibinfo {author} {\bibfnamefont {W.~M.}\ \bibnamefont
  {Itano}}, \ and\ \bibinfo {author} {\bibfnamefont {D.~J.}\ \bibnamefont
  {Wineland}},\ }\href {\doibase 10.1103/PhysRevLett.57.70} {\bibfield
  {journal} {\bibinfo  {journal} {Phys. Rev. Lett.}\ }\textbf {\bibinfo
  {volume} {57}},\ \bibinfo {pages} {70} (\bibinfo {year} {1986})}\BibitemShut
  {NoStop}%
\bibitem [{\citenamefont {Lee}\ and\ \citenamefont
  {Cary}(2005)}]{PhysRevE.71.036406}%
  \BibitemOpen
  \bibfield  {author} {\bibinfo {author} {\bibfnamefont {J.}~\bibnamefont
  {Lee}}\ and\ \bibinfo {author} {\bibfnamefont {J.~R.}\ \bibnamefont {Cary}},\
  }\href {\doibase 10.1103/PhysRevE.71.036406} {\bibfield  {journal} {\bibinfo
  {journal} {Phys. Rev. E}\ }\textbf {\bibinfo {volume} {71}},\ \bibinfo
  {pages} {036406} (\bibinfo {year} {2005})}\BibitemShut {NoStop}%
\bibitem [{\citenamefont {Surko}\ and\ \citenamefont
  {Greaves}(2004)}]{surko2004emerging}%
  \BibitemOpen
  \bibfield  {author} {\bibinfo {author} {\bibfnamefont {C.~M.}\ \bibnamefont
  {Surko}}\ and\ \bibinfo {author} {\bibfnamefont {R.~G.}\ \bibnamefont
  {Greaves}},\ }\href {\doibase 10.1063/1.1651487} {\bibfield  {journal}
  {\bibinfo  {journal} {Phys. Plasmas}\ }\textbf {\bibinfo {volume} {11}},\
  \bibinfo {pages} {2333} (\bibinfo {year} {2004})}\BibitemShut {NoStop}%
\bibitem [{\citenamefont {Folegati}\ \emph {et~al.}(2011)\citenamefont
  {Folegati}, \citenamefont {Xu}, \citenamefont {Weber},\ and\ \citenamefont
  {Lynn}}]{folegati2011positron}%
  \BibitemOpen
  \bibfield  {author} {\bibinfo {author} {\bibfnamefont {P.}~\bibnamefont
  {Folegati}}, \bibinfo {author} {\bibfnamefont {J.}~\bibnamefont {Xu}},
  \bibinfo {author} {\bibfnamefont {M.~H.}\ \bibnamefont {Weber}}, \ and\
  \bibinfo {author} {\bibfnamefont {K.~G.}\ \bibnamefont {Lynn}},\ }in\
  \href@noop {} {\emph {\bibinfo {booktitle} {J. Phys. Conf. Ser.}}},\ Vol.\
  \bibinfo {volume} {262}\ (\bibinfo {organization} {IOP Publishing},\ \bibinfo
  {year} {2011})\ p.\ \bibinfo {pages} {012021}\BibitemShut {NoStop}%
\bibitem [{\citenamefont {Boyer}(1974)}]{boyer1974penetration}%
  \BibitemOpen
  \bibfield  {author} {\bibinfo {author} {\bibfnamefont {T.~H.}\ \bibnamefont
  {Boyer}},\ }\href {\doibase 10.1103/PhysRevA.9.68} {\bibfield  {journal}
  {\bibinfo  {journal} {Phys. Rev. A}\ }\textbf {\bibinfo {volume} {9}},\
  \bibinfo {pages} {68} (\bibinfo {year} {1974})}\BibitemShut {NoStop}%
\bibitem [{\citenamefont {Mollenstedt}\ and\ \citenamefont
  {Bayh}(1962)}]{Mollenstedt}%
  \BibitemOpen
  \bibfield  {author} {\bibinfo {author} {\bibfnamefont {G.}~\bibnamefont
  {Mollenstedt}}\ and\ \bibinfo {author} {\bibfnamefont {W.}~\bibnamefont
  {Bayh}},\ }\href@noop {} {\bibfield  {journal} {\bibinfo  {journal} {Z. Phys.
  A-Hadron. Nucl.}\ }\textbf {\bibinfo {volume} {169}},\ \bibinfo {pages} {492}
  (\bibinfo {year} {1962})}\BibitemShut {NoStop}%
\bibitem [{\citenamefont {Kasper}(1966)}]{Ekasper1966}%
  \BibitemOpen
  \bibfield  {author} {\bibinfo {author} {\bibfnamefont {E.}~\bibnamefont
  {Kasper}},\ }\href@noop {} {\bibfield  {journal} {\bibinfo  {journal} {Z.
  Phys. A-Hadron. Nucl.}\ }\textbf {\bibinfo {volume} {196}},\ \bibinfo {pages}
  {415} (\bibinfo {year} {1966})}\BibitemShut {NoStop}%
\bibitem [{\citenamefont {Boyer}(1999)}]{boyer1999understanding}%
  \BibitemOpen
  \bibfield  {author} {\bibinfo {author} {\bibfnamefont {T.~H.}\ \bibnamefont
  {Boyer}},\ }\href {\doibase 10.1119/1.19171} {\bibfield  {journal} {\bibinfo
  {journal} {Am. J. of Phys.}\ }\textbf {\bibinfo {volume} {67}},\ \bibinfo
  {pages} {954} (\bibinfo {year} {1999})}\BibitemShut {NoStop}%
\bibitem [{\citenamefont {Boyer}(1996)}]{boyer1996penetration}%
  \BibitemOpen
  \bibfield  {author} {\bibinfo {author} {\bibfnamefont {T.~H.}\ \bibnamefont
  {Boyer}},\ }\href@noop {} {\bibfield  {journal} {\bibinfo  {journal} {Phys.
  Rev. E}\ }\textbf {\bibinfo {volume} {53}},\ \bibinfo {pages} {6450}
  (\bibinfo {year} {1996})}\BibitemShut {NoStop}%
\bibitem [{\citenamefont {Jones}(1975)}]{0305-4470-8-5-009}%
  \BibitemOpen
  \bibfield  {author} {\bibinfo {author} {\bibfnamefont {D.~S.}\ \bibnamefont
  {Jones}},\ }\href {http://stacks.iop.org/0305-4470/8/i=5/a=009} {\bibfield
  {journal} {\bibinfo  {journal} {Journal of Physics A: Mathematical and
  General}\ }\textbf {\bibinfo {volume} {8}},\ \bibinfo {pages} {742} (\bibinfo
  {year} {1975})}\BibitemShut {NoStop}%
\bibitem [{\citenamefont {Aguirregabiria}\ \emph {et~al.}(1995)\citenamefont
  {Aguirregabiria}, \citenamefont {Hernández},\ and\ \citenamefont
  {Rivas}}]{Aguirregabiria19956}%
  \BibitemOpen
  \bibfield  {author} {\bibinfo {author} {\bibfnamefont {J.}~\bibnamefont
  {Aguirregabiria}}, \bibinfo {author} {\bibfnamefont {A.}~\bibnamefont
  {Hernández}}, \ and\ \bibinfo {author} {\bibfnamefont {M.}~\bibnamefont
  {Rivas}},\ }\href {\doibase http://dx.doi.org/10.1016/0375-9601(94)01018-P}
  {\bibfield  {journal} {\bibinfo  {journal} {Physics Letters A}\ }\textbf
  {\bibinfo {volume} {198}},\ \bibinfo {pages} {6 } (\bibinfo {year}
  {1995})}\BibitemShut {NoStop}%
\bibitem [{\citenamefont {Schaich}(2001)}]{PhysRevE.64.046605}%
  \BibitemOpen
  \bibfield  {author} {\bibinfo {author} {\bibfnamefont {W.~L.}\ \bibnamefont
  {Schaich}},\ }\href {\doibase 10.1103/PhysRevE.64.046605} {\bibfield
  {journal} {\bibinfo  {journal} {Phys. Rev. E}\ }\textbf {\bibinfo {volume}
  {64}},\ \bibinfo {pages} {046605} (\bibinfo {year} {2001})}\BibitemShut
  {NoStop}%
\bibitem [{\citenamefont {Kauppila}\ and\ \citenamefont
  {Stein}(1989)}]{Kauppila19891}%
  \BibitemOpen
  \bibfield  {author} {\bibinfo {author} {\bibfnamefont {W.~E.}\ \bibnamefont
  {Kauppila}}\ and\ \bibinfo {author} {\bibfnamefont {T.~S.}\ \bibnamefont
  {Stein}},\ }\href@noop {} {\bibfield  {journal} {\bibinfo  {journal} {Adv.
  Atom. Mol. Opt. Phys.}\ }\textbf {\bibinfo {volume} {26}},\ \bibinfo {pages}
  {1} (\bibinfo {year} {1989})}\BibitemShut {NoStop}%
\bibitem [{\citenamefont {Greaves}\ \emph {et~al.}(1994)\citenamefont
  {Greaves}, \citenamefont {Tinkle},\ and\ \citenamefont
  {Surko}}]{greaves:1439}%
  \BibitemOpen
  \bibfield  {author} {\bibinfo {author} {\bibfnamefont {R.~G.}\ \bibnamefont
  {Greaves}}, \bibinfo {author} {\bibfnamefont {M.~D.}\ \bibnamefont {Tinkle}},
  \ and\ \bibinfo {author} {\bibfnamefont {C.~M.}\ \bibnamefont {Surko}},\
  }\href {\doibase 10.1063/1.870693} {\bibfield  {journal} {\bibinfo  {journal}
  {Phys. Plasmas}\ }\textbf {\bibinfo {volume} {1}},\ \bibinfo {pages} {1439}
  (\bibinfo {year} {1994})}\BibitemShut {NoStop}%
\bibitem [{\citenamefont {Brown}\ \emph {et~al.}(1984)\citenamefont {Brown},
  \citenamefont {Leventhal}, \citenamefont {Mills},\ and\ \citenamefont
  {Gidley}}]{PhysRevLett.53.2347}%
  \BibitemOpen
  \bibfield  {author} {\bibinfo {author} {\bibfnamefont {B.~L.}\ \bibnamefont
  {Brown}}, \bibinfo {author} {\bibfnamefont {M.}~\bibnamefont {Leventhal}},
  \bibinfo {author} {\bibfnamefont {A.~P.}\ \bibnamefont {Mills}}, \ and\
  \bibinfo {author} {\bibfnamefont {D.~W.}\ \bibnamefont {Gidley}},\ }\href
  {\doibase 10.1103/PhysRevLett.53.2347} {\bibfield  {journal} {\bibinfo
  {journal} {Phys. Rev. Lett.}\ }\textbf {\bibinfo {volume} {53}},\ \bibinfo
  {pages} {2347} (\bibinfo {year} {1984})}\BibitemShut {NoStop}%
\bibitem [{\citenamefont {Jr.}\ \emph {et~al.}(1993)\citenamefont {Jr.},
  \citenamefont {Donohue}, \citenamefont {Xu}, \citenamefont {Lewis},
  \citenamefont {McLuckey},\ and\ \citenamefont {Glish}}]{HulettJr1993236}%
  \BibitemOpen
  \bibfield  {author} {\bibinfo {author} {\bibfnamefont {L.~H.}\ \bibnamefont
  {Jr.}}, \bibinfo {author} {\bibfnamefont {D.}~\bibnamefont {Donohue}},
  \bibinfo {author} {\bibfnamefont {J.}~\bibnamefont {Xu}}, \bibinfo {author}
  {\bibfnamefont {T.}~\bibnamefont {Lewis}}, \bibinfo {author} {\bibfnamefont
  {S.}~\bibnamefont {McLuckey}}, \ and\ \bibinfo {author} {\bibfnamefont
  {G.}~\bibnamefont {Glish}},\ }\href {\doibase 10.1016/0009-2614(93)E1231-5}
  {\bibfield  {journal} {\bibinfo  {journal} {Chem. Phys. Lett.}\ }\textbf
  {\bibinfo {volume} {216}},\ \bibinfo {pages} {236 } (\bibinfo {year}
  {1993})}\BibitemShut {NoStop}%
\bibitem [{\citenamefont {Charlton}\ \emph {et~al.}(1994)\citenamefont
  {Charlton}, \citenamefont {Eades}, \citenamefont {HorvÃ¡th}, \citenamefont
  {Hughes},\ and\ \citenamefont {Zimmermann}}]{Charlton199465}%
  \BibitemOpen
  \bibfield  {author} {\bibinfo {author} {\bibfnamefont {M.}~\bibnamefont
  {Charlton}}, \bibinfo {author} {\bibfnamefont {J.}~\bibnamefont {Eades}},
  \bibinfo {author} {\bibfnamefont {D.}~\bibnamefont {HorvÃ¡th}}, \bibinfo
  {author} {\bibfnamefont {R.}~\bibnamefont {Hughes}}, \ and\ \bibinfo {author}
  {\bibfnamefont {C.}~\bibnamefont {Zimmermann}},\ }\href {\doibase
  10.1016/0370-1573(94)90081-7} {\bibfield  {journal} {\bibinfo  {journal}
  {Phys. Rep.}\ }\textbf {\bibinfo {volume} {241}},\ \bibinfo {pages} {65 }
  (\bibinfo {year} {1994})}\BibitemShut {NoStop}%
\bibitem [{\citenamefont {Krause-Rehberg}\ and\ \citenamefont
  {Leipner}(1999)}]{Krause-Rehberg1999positron}%
  \BibitemOpen
  \bibfield  {author} {\bibinfo {author} {\bibfnamefont {R.}~\bibnamefont
  {Krause-Rehberg}}\ and\ \bibinfo {author} {\bibfnamefont {H.~S.}\
  \bibnamefont {Leipner}},\ }\href@noop {} {\emph {\bibinfo {title} {Positron
  annihilation in semiconductors: defect studies}}}\ (\bibinfo  {publisher}
  {Springer},\ \bibinfo {year} {1999})\BibitemShut {NoStop}%
\bibitem [{\citenamefont {Schultz}\ and\ \citenamefont
  {Lynn}(1988)}]{lynn1988interaction}%
  \BibitemOpen
  \bibfield  {author} {\bibinfo {author} {\bibfnamefont {P.~J.}\ \bibnamefont
  {Schultz}}\ and\ \bibinfo {author} {\bibfnamefont {K.~G.}\ \bibnamefont
  {Lynn}},\ }\href {\doibase 10.1103/RevModPhys.60.701} {\bibfield  {journal}
  {\bibinfo  {journal} {Rev. Mod. Phys.}\ }\textbf {\bibinfo {volume} {60}},\
  \bibinfo {pages} {701} (\bibinfo {year} {1988})}\BibitemShut {NoStop}%
\bibitem [{\citenamefont {Surko}\ and\ \citenamefont
  {Greaves}(2003)}]{surko2003multicell}%
  \BibitemOpen
  \bibfield  {author} {\bibinfo {author} {\bibfnamefont {C.~M.}\ \bibnamefont
  {Surko}}\ and\ \bibinfo {author} {\bibfnamefont {R.~G.}\ \bibnamefont
  {Greaves}},\ }\href@noop {} {\bibfield  {journal} {\bibinfo  {journal}
  {Radiat. Phys. Chem.}\ }\textbf {\bibinfo {volume} {68}},\ \bibinfo {pages}
  {419} (\bibinfo {year} {2003})}\BibitemShut {NoStop}%
\bibitem [{\citenamefont {Danielson}\ \emph {et~al.}(2006)\citenamefont
  {Danielson}, \citenamefont {Weber},\ and\ \citenamefont
  {Surko}}]{danielson2006plasma}%
  \BibitemOpen
  \bibfield  {author} {\bibinfo {author} {\bibfnamefont {J.~R.}\ \bibnamefont
  {Danielson}}, \bibinfo {author} {\bibfnamefont {T.~R.}\ \bibnamefont
  {Weber}}, \ and\ \bibinfo {author} {\bibfnamefont {C.~M.}\ \bibnamefont
  {Surko}},\ }\href@noop {} {\bibfield  {journal} {\bibinfo  {journal} {Phys.
  plasmas}\ }\textbf {\bibinfo {volume} {13}},\ \bibinfo {pages} {123502}
  (\bibinfo {year} {2006})}\BibitemShut {NoStop}%
\bibitem [{\citenamefont {Greaves}\ and\ \citenamefont
  {Surko}(2002)}]{greaves2002practical}%
  \BibitemOpen
  \bibfield  {author} {\bibinfo {author} {\bibfnamefont {R.~G.}\ \bibnamefont
  {Greaves}}\ and\ \bibinfo {author} {\bibfnamefont {C.~M.}\ \bibnamefont
  {Surko}},\ }\href@noop {} {\emph {\bibinfo {title} {Practical limits on
  positron accumulation and the creation of electron-positron plasmas}}},\
  \bibinfo {type} {Tech. Rep.}\ (\bibinfo  {institution} {DTIC Document},\
  \bibinfo {year} {2002})\BibitemShut {NoStop}%
\bibitem [{\citenamefont {Lynn}\ and\ \citenamefont
  {Greaves}(2001)}]{lynngreaves2001longaspect}%
  \BibitemOpen
  \bibfield  {author} {\bibinfo {author} {\bibfnamefont {K.~G.}\ \bibnamefont
  {Lynn}}\ and\ \bibinfo {author} {\bibfnamefont {R.~G.}\ \bibnamefont
  {Greaves}},\ }\href@noop {} {\enquote {\bibinfo {title} {Long aspect ratio
  positron micro-trap},}\ } (\bibinfo {year} {2001}),\ \bibinfo {note} {private
  communication}\BibitemShut {NoStop}%
\bibitem [{\citenamefont {Narimannezhad}\ \emph {et~al.}(2012)\citenamefont
  {Narimannezhad}, \citenamefont {Xu}, \citenamefont {Baker}, \citenamefont
  {Weber},\ and\ \citenamefont {Lynn}}]{jialireza2012feasibility}%
  \BibitemOpen
  \bibfield  {author} {\bibinfo {author} {\bibfnamefont {A.}~\bibnamefont
  {Narimannezhad}}, \bibinfo {author} {\bibfnamefont {J.}~\bibnamefont {Xu}},
  \bibinfo {author} {\bibfnamefont {C.~J.}\ \bibnamefont {Baker}}, \bibinfo
  {author} {\bibfnamefont {M.~H.}\ \bibnamefont {Weber}}, \ and\ \bibinfo
  {author} {\bibfnamefont {K.~G.}\ \bibnamefont {Lynn}},\ }\href@noop {}
  {\enquote {\bibinfo {title} {Simulation studies of positrons bahaviour in a
  microtrap with long aspect ratio},}\ } (\bibinfo {year} {2012}),\ \Eprint
  {http://arxiv.org/abs/1301.0030v1} {arXiv:1301.0030v1} \BibitemShut {NoStop}%
\bibitem [{\citenamefont {Jackson}(1998)}]{jackson1998classical}%
  \BibitemOpen
  \bibfield  {author} {\bibinfo {author} {\bibfnamefont {J.~D.}\ \bibnamefont
  {Jackson}},\ }\enquote {\bibinfo {title} {Classical electrodynamics},}\ \
  (\bibinfo  {publisher} {John Wiley \& Sons},\ \bibinfo {year} {1998})\ p.\
  \bibinfo {pages} {143}\BibitemShut {NoStop}%
\bibitem [{\citenamefont {Clarke}\ \emph {et~al.}(2006)\citenamefont {Clarke},
  \citenamefont {van~der Werf}, \citenamefont {Griffiths}, \citenamefont
  {Beddows}, \citenamefont {Charlton}, \citenamefont {Telle},\ and\
  \citenamefont {Watkeys}}]{clarke2006design}%
  \BibitemOpen
  \bibfield  {author} {\bibinfo {author} {\bibfnamefont {J.}~\bibnamefont
  {Clarke}}, \bibinfo {author} {\bibfnamefont {D.~P.}\ \bibnamefont {van~der
  Werf}}, \bibinfo {author} {\bibfnamefont {B.}~\bibnamefont {Griffiths}},
  \bibinfo {author} {\bibfnamefont {D.~C.~S.}\ \bibnamefont {Beddows}},
  \bibinfo {author} {\bibfnamefont {M.}~\bibnamefont {Charlton}}, \bibinfo
  {author} {\bibfnamefont {H.~H.}\ \bibnamefont {Telle}}, \ and\ \bibinfo
  {author} {\bibfnamefont {P.~R.}\ \bibnamefont {Watkeys}},\ }\href {\doibase
  10.1063/1.2206561} {\bibfield  {journal} {\bibinfo  {journal} {Rev. of Sci.
  Instrum.}\ }\textbf {\bibinfo {volume} {77}},\ \bibinfo {eid} {063302}
  (\bibinfo {year} {2006})}\BibitemShut {NoStop}%
\bibitem [{\citenamefont {Al-Qaradawi}\ \emph {et~al.}(2000)\citenamefont
  {Al-Qaradawi}, \citenamefont {Charlton}, \citenamefont {Borozan},
  \citenamefont {Whitehead},\ and\ \citenamefont
  {Borozan}}]{Ilham2000thermalization}%
  \BibitemOpen
  \bibfield  {author} {\bibinfo {author} {\bibfnamefont {I.}~\bibnamefont
  {Al-Qaradawi}}, \bibinfo {author} {\bibfnamefont {M.}~\bibnamefont
  {Charlton}}, \bibinfo {author} {\bibfnamefont {I.}~\bibnamefont {Borozan}},
  \bibinfo {author} {\bibfnamefont {R.}~\bibnamefont {Whitehead}}, \ and\
  \bibinfo {author} {\bibfnamefont {I.}~\bibnamefont {Borozan}},\ }\href
  {http://stacks.iop.org/0953-4075/33/i=14/a=309} {\bibfield  {journal}
  {\bibinfo  {journal} {J. Phys. B-At. Mol. Opt.}\ }\textbf {\bibinfo {volume}
  {33}},\ \bibinfo {pages} {2725} (\bibinfo {year} {2000})}\BibitemShut
  {NoStop}%
\bibitem [{\citenamefont {Murphy}\ and\ \citenamefont
  {Surko}(1992)}]{PhysRevA.46.5696}%
  \BibitemOpen
  \bibfield  {author} {\bibinfo {author} {\bibfnamefont {T.~J.}\ \bibnamefont
  {Murphy}}\ and\ \bibinfo {author} {\bibfnamefont {C.~M.}\ \bibnamefont
  {Surko}},\ }\href {\doibase 10.1103/PhysRevA.46.5696} {\bibfield  {journal}
  {\bibinfo  {journal} {Phys. Rev. A}\ }\textbf {\bibinfo {volume} {46}},\
  \bibinfo {pages} {5696} (\bibinfo {year} {1992})}\BibitemShut {NoStop}%
\bibitem [{\citenamefont {Beck}\ \emph {et~al.}(1996)\citenamefont {Beck},
  \citenamefont {Fajans},\ and\ \citenamefont {Malmberg}}]{beck:1250}%
  \BibitemOpen
  \bibfield  {author} {\bibinfo {author} {\bibfnamefont {B.~R.}\ \bibnamefont
  {Beck}}, \bibinfo {author} {\bibfnamefont {J.}~\bibnamefont {Fajans}}, \ and\
  \bibinfo {author} {\bibfnamefont {J.~H.}\ \bibnamefont {Malmberg}},\ }\href
  {\doibase 10.1063/1.871749} {\bibfield  {journal} {\bibinfo  {journal}
  {Physics of Plasmas}\ }\textbf {\bibinfo {volume} {3}},\ \bibinfo {pages}
  {1250} (\bibinfo {year} {1996})}\BibitemShut {NoStop}%
\bibitem [{\citenamefont {Beck}(1990)}]{BentPhd05}%
  \BibitemOpen
  \bibfield  {author} {\bibinfo {author} {\bibfnamefont {B.~R.}\ \bibnamefont
  {Beck}},\ }\href@noop {} {Ph.D. thesis},\ \bibinfo  {school} {University of
  California, San Diego} (\bibinfo {year} {1990})\BibitemShut {NoStop}%
\bibitem [{\citenamefont {Lourtioz}\ \emph {et~al.}(2008)\citenamefont
  {Lourtioz}, \citenamefont {Benisty}, \citenamefont {Pagnoux}, \citenamefont
  {Berger}, \citenamefont {Gerard},\ and\ \citenamefont
  {Maystre}}]{lourtioz2008photonic}%
  \BibitemOpen
  \bibfield  {author} {\bibinfo {author} {\bibfnamefont {J.~M.}\ \bibnamefont
  {Lourtioz}}, \bibinfo {author} {\bibfnamefont {H.}~\bibnamefont {Benisty}},
  \bibinfo {author} {\bibfnamefont {D.}~\bibnamefont {Pagnoux}}, \bibinfo
  {author} {\bibfnamefont {V.}~\bibnamefont {Berger}}, \bibinfo {author}
  {\bibfnamefont {J.~M.}\ \bibnamefont {Gerard}}, \ and\ \bibinfo {author}
  {\bibfnamefont {D.}~\bibnamefont {Maystre}},\ }\enquote {\bibinfo {title}
  {Photonic crystals: towards nanoscale photonic devices},}\ \ (\bibinfo {year}
  {2008})\ p.\ \bibinfo {pages} {122}\BibitemShut {NoStop}%
\end{thebibliography}%

\end{document}